\documentclass[11pt]{article}

\usepackage[margin=1in]{geometry}
\usepackage{microtype}
\usepackage{graphicx}
\usepackage{booktabs}
\usepackage{makecell}
\usepackage{multirow}
\usepackage{amsmath}
\usepackage{amssymb}
\usepackage{pdflscape}
\usepackage[numbers,sort&compress]{natbib}
\usepackage{placeins}
\usepackage[hidelinks]{hyperref}

\title{AI-written admissions essays are widespread but penalized}

\author{
  Calvin Isley\\
  Harvard University
  \and
  Johann D. Gaebler\\
  New York University
  \and
  Sharad Goel\\
  Harvard University
}

\date{}

\begin{document}

\maketitle

\begin{abstract}
AI is rapidly transforming higher education, including the application process, yet relatively little is known about its use and consequences. To help close this gap, we analyze nearly 7{,}500 applications submitted between 2020 and 2025 to a large public policy master's program in the United States. We find that in the 2025 admissions cycle, the majority of applicants submitted at least one essay that was primarily AI-generated---despite an explicit prohibition against using AI. Leveraging the abrupt introduction of ChatGPT in November 2022, we find that the availability of AI assistants improved the writing quality of submitted essays. These improvements, however, came with an apparent AI penalty: Applicants submitting AI-written essays were admitted less often than comparable non-users. To help explain this penalty, we conduct an experiment with admissions officers, finding that they can often recognize AI writing and rate essays they believe to be AI-generated lower than essays they believe to be human generated. These findings indicate that AI is changing both how applicants write and how that writing is evaluated, raising questions about whether admissions practices and policies designed for a pre-AI era remain appropriate.
\end{abstract}
\vfill
\paragraph{Acknowledgments} We thank the admissions staff at our partner institution for data access and guidance, GPTZero for providing free API access, and Alex Adam for helpful conversations.

\newpage

\section{Introduction}
When ChatGPT was released in November 2022~\citep{noauthor_introducing_2024}, it fundamentally altered the way people write. AI-generated writing has since proliferated across a variety of domains, ranging from academic journal submissions, to job applications, to UN press briefings~\citep{liang_quantifying_2025, kessler_employers_2025, spennemann_delving_2025,russell_ai_2026,doi:10.1287/orsc.2026.ed.v37.n3}. University admissions essays have been no exception~\citep{singer_ban_2023, moon_creative_2026, digitaldividelee2026}. Historically, admissions essays have been a channel for applicants to demonstrate their writing ability, motivation, and program fit, beyond grades and test scores~\citep{ross_what_2020, bastedo_what_2018, rigol_admissions_2003}. Now, however, university admissions offices must contend with the possibility that applicants' submitted essays do not reflect their abilities or authentic voice~\citep{singer_ban_2023}. Recent survey evidence suggests that AI use may already be widespread, with 30\% of college applicants reporting using generative AI to help write their personal essays, 20\% of whom reported using AI to produce a final draft~\citep{rubin_navigating_2024}. Similarly, Lee et al.~\citep{digitaldividelee2026} estimated that approximately 8–10\% of the text in Common Application essays submitted to a competitive engineering school exhibited linguistic patterns characteristic of LLM-generated content in the year after ChatGPT's release.

Despite the speed, scale, and significance of this change, few universities provide formal guidelines on the use of AI in applications~\citep{kaplan_kaplan_2025}, perhaps reflecting a lack of evidence on the prevalence and consequences of AI usage in admissions. Recent work has considered the detectability of AI in admissions materials~\citep{zhao_admissions_2024, burke_artificial_2025, cumbo_can_nodate, lum_can_2024}, the stylistic properties of AI-generated admissions essays~\citep{digitaldividelee2026, lee2026pooralignmentsteerabilitylarge, moon_creative_2026}, and exploratory analysis of the impacts of AI use on admissions outcomes~\citep{digitaldividelee2026}. But it remains unclear how often applicants are using AI, how AI impacts the quality of their applications, and whether AI use affects applicants' admission decisions. Without answers to these questions, admissions offices may struggle to set and enforce sensible AI policies. This lack of clarity leaves applicants to make individual choices on the appropriate use of AI, potentially exacerbating disparities in the admissions process.

Here we begin to close this information gap by analyzing a novel dataset of nearly 7{,}500 applications to a large American public policy master's program submitted between 2020 and 2025---covering three admissions cycles before and three after the release of ChatGPT. Using commercial AI detection tools, we find that AI usage is widespread and growing, with 56\% of applicants submitting at least one essay that was likely primarily AI-generated in the most recent application cycle. Comparing pre- and post-ChatGPT essays, we find that the broad availability of AI substantially improved the quality of applicants' essays, particularly for international applicants.

Despite these improvements in essay quality, applicants who submitted AI-written essays were nevertheless admitted at lower rates than otherwise similar applicants who did not use AI. To understand the source of this apparent penalty, we conduct an experiment to measure admissions officers' ability to detect AI-written essays. We find that not only can admissions officers often discriminate between AI and human-written text, but that they also rate essays they suspect to be AI-written more harshly. Together, our findings suggest that AI use is both common and consequential, at least among the applicants we study, but likely more broadly. These results underscore the need for universities to reevaluate the admissions process in response to this new reality, and to develop sensible and enforceable AI policies for it. 

\section{Data}

To study the role of AI writing in admissions, we assemble a novel dataset of 7{,}462 applications to a competitive U.S. public policy master's program submitted across six admissions cycles between 2020 and 2025. In addition to five admissions essays, prospective students provide basic demographic information, GRE or GMAT scores,\footnote{%
    Applicants were not required to submit GRE or GMAT scores if they applied in 2020 or 2021.
}
academic transcripts from previous undergraduate or graduate institutions, and three letters of recommendation from university faculty members or immediate professional supervisors. The essays ranged in length from 250 to 500 words and cover topics ranging from applicants' motivation for applying to the program, to their personal and professional background, to their plans to make an impact through public service.\footnote{%
    The prompts for the required essays were largely the same throughout our sample period, barring a substantive change to one essay's prompt in 2023. Our results persist regardless of whether we include or exclude that essay. See Appendix Section~\ref{sapp:robustness} for discussion of the modified results.
} Applicants who did not attend an English-language undergraduate institution were also required to submit TOEFL, IELTS or Cambridge English scores. These materials comprise the totality of information about applicants available to admissions officers during their decision-making process. 

To facilitate statistical analysis, we designed an LLM-based pipeline to convert the unstructured application materials to structured measures~\citep{gaebler_test_scores,isley2026mitigatinglabelbiasinterpretable}. These measures include, for example: length of prior public service or non-profit work experience; performance in previous program-relevant coursework, standardized to a 4.0 scale; the number and severity of grammatical errors in essays; and the strength of endorsement in submitted recommendation letters. In total, we extract 216 such measures from the unstructured materials. These measures were designed in collaboration with admissions officers in the program we analyze, with the aim of capturing the key information informing admissions decisions. (For full details, see Appendix Section~\ref{sapp:data}.)

\section{Results}

For all six admissions cycles in our data, the program we study required applications to be completed personally and in English. In particular, before final submission, applicants signed an attestation that they did not receive prohibited writing or translation assistance. Starting in 2023---the first cycle after ChatGPT was released---this attestation was expanded to exclude use of generative AI.

\subsection{AI-written essays are common}

\begin{figure}[t]
\centering
\includegraphics{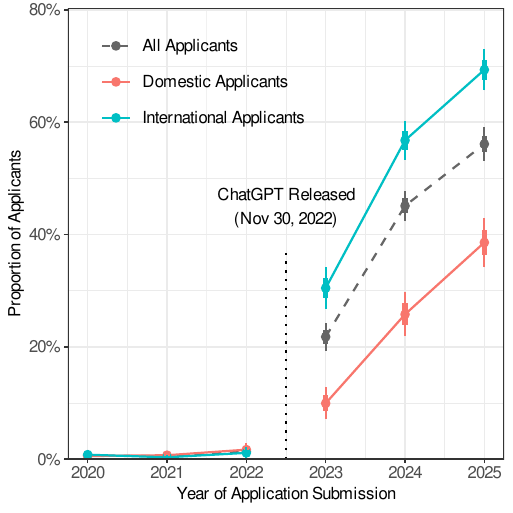}
\caption{%
    \emph{Proportion of applicants who submitted one or more essays detected by GPTZero as written by AI between 2020 and 2025. Solid lines are disaggregated by nationality, the dashed line aggregates across all applicants. ChatGPT was released on Nov.~30, 2022~\citep{noauthor_introducing_2024}, one day before the December 1 application deadline for the 2022 cycle, suggesting few, if any, applications were written with AI by applicants that year.}
    }
\label{fig:timeseries}
\end{figure}

We use a popular commercial AI-detection tool, GPTZero~\citep{adam2026gptzerorobustdetectionllmgenerated}, to identify AI-generated essays. GPTZero and other tools like it work by identifying subtle stylistic patterns in text to determine its origin. GPTZero classifies documents into one of three categories: (1) ``AI written,'' meaning documents primarily produced by AI, including human-written text that was rewritten by AI; (2) ``human,'' meaning documents written exclusively by humans; and (3) ``mixed,'' meaning documents containing both human-written and AI-generated text. GPTZero first estimates the probability that text comes from each of these three categories, and then assigns documents to the highest-probability category. It further assigns confidence scores to the classifications indicating their level of certainty. For our primary analysis, we use its ``AI written'' classification with ``high'' confidence, suggesting that flagged essays were primarily generated by AI---and we throughout refer to such essays to be ``AI-generated.'' We replicate our primary results with Pangram, as well as with looser AI-use classification thresholds (e.g., including documents classified as ``mixed''), finding qualitatively similar results. See Appendix~\ref{sapp:robustness} for details.

For each of the six application cycles we study, Figure~\ref{fig:timeseries} shows the proportion of applicants flagged as submitting at least one AI-generated essay. In the three cycles prior to the release of ChatGPT in late 2022, virtually no essays were identified as AI-written. That pattern suggests the detector has a low false-positive rate, consistent with the company's stated error rate as well as with estimates from independent audits~\citep{jabarian_artificial_2025, adam2026gptzerorobustdetectionllmgenerated}.

Beginning in 2023, the proportion of applicants submitting at least one essay flagged as AI-written climbs by around \(20\,\mathrm{p.p.}\) per year on average, to nearly 60\% in the most recent cycle. We find that AI usage is also considerably higher for international applicants (69.3\%) than for domestic applicants (38.6\%).\footnote{%
    In contrast to prior work on the quality of AI detection software, we find no evidence of a differential false positive rate for international applicants or for TOEFL and IELTS takers~\citep{liang_gpt_2023}. The usage gap between domestic and international applicants is larger than the gap between applicants who submit language test scores and those who do not.
}
We further find that AI use is higher among younger applicants, but is mostly uncorrelated with performance on standardized tests and undergraduate major (see Appendix Fig.~\ref{fig:usage_by_demo}).

Though already relatively high, our estimate likely understates the true prevalence of AI use. AI detectors are typically calibrated to ensure few human-written responses are erroneously flagged as AI-generated, consistent with the low false-positive rate we see in the pre-ChatGPT years in Figure~\ref{fig:timeseries}. As a result, however, these detectors could have relatively high false-negative rates, mistakenly classifying AI-generated text as human. Further, to aid interpretation, we have focused on text that was likely primarily generated by AI. As a result, our counts miss instances where applicants use AI in part without submitting fully AI-generated text, behavior that is still prohibited by the program we consider.

\subsection{AI improves essay quality}

\begin{figure}
\centering
\includegraphics{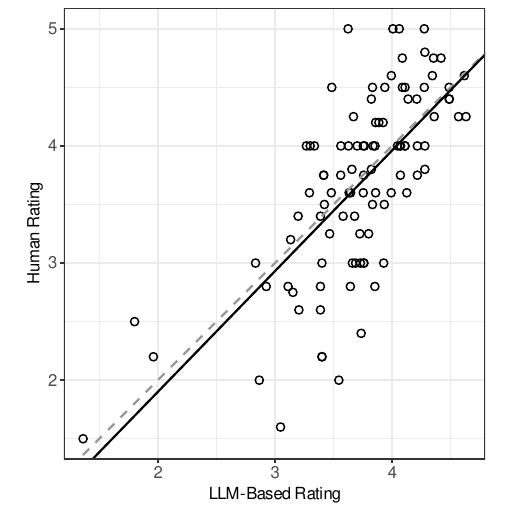}
\caption{%
    \emph{Comparison of expert and LLM-based ratings of essay quality. For each essay, its expert rating is the average rating given by admissions officers, on a six-point scale; the LLM-based rating is a weighted combination of seven dimensions of essay quality, scored by an LLM. Higher values indicate higher quality. The solid line is the line of best fit, and the dashed line indicates equality between human and model scores. The two rating methods have a correlation of 70\%.}}
\label{fig:human_model_calib}
\end{figure}

We next consider the impact of AI on the quality of submitted essays. To do so, we develop an automated procedure to evaluate essays along seven dimensions: grammar, style, clarity, prompt responsiveness, voice, commitment to public service, and demonstrated leadership. We instruct an LLM to grade each dimension on a five-point scale designed to minimize interpretive ambiguity. For example, each essay's grammar score reflects the number and severity of errors, with 5 points awarded for 0--1 surface errors (e.g., spelling, punctuation, and subject-verb misalignment), and 1 point for either 11+ such errors or multiple meaning-blocking errors. To construct a single, composite score of essay quality, we then had five admissions officers from the partner program each rate up to 100 submitted essays---50 that were likely generated by humans and 50 by AI---on a six-point scale, from ``much weaker'' to ``much stronger'' than a typical essay. Finally, we fit a model to predict these expert ratings from the LLM-scored measures, yielding weights for each of the seven dimensions of essay quality we consider---and, accordingly, a composite essay quality score. Figure~\ref{fig:human_model_calib} shows that the automated and expert ratings are in relatively close agreement, with a correlation of 70\%. Further details are provided in Appendix Sections~\ref{sapp:experiment} and~\ref{sapp:essay_quality}.

\begin{figure*}
\centering
\includegraphics{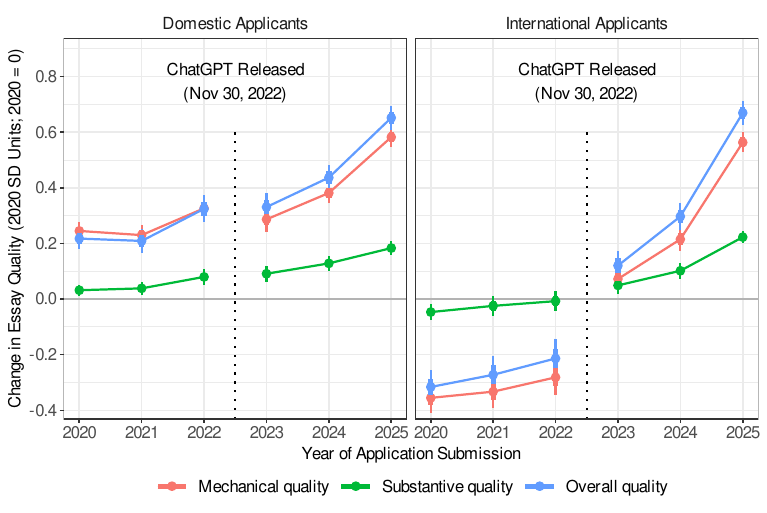}
\caption{%
    \emph{Change in human-calibrated and LLM-assessed essay quality over time, disaggregated by residency status, in 2020 standard deviation units. LLM quality ratings are aggregated into substantive quality and mechanical quality. Each series is independently standardized to the mean and standard deviation of the full applicant pool applying in 2020.}
    } 
    \label{fig:quality_over_time}
\end{figure*}

We apply our automated evaluation procedure to infer overall and dimension-specific quality scores for each essay in our dataset. Then, for each applicant, we average the scores of their submitted essays to produce applicant-level measures of writing quality. Figure~\ref{fig:quality_over_time} shows a sharp improvement in the overall quality of essays (blue line) after the introduction of ChatGPT, particularly for international applicants and especially in the most recent cycles, when use of AI was greatest. These gains come from large improvements in mechanical aspects of writing (grammar, style, and clarity, indicated by the red line), with more modest improvement along substantive dimensions (prompt responsiveness, voice, commitment to public service, and demonstrated leadership, indicated by the green line). Further, these overall and dimension-specific gains in essay quality persist when we include controls for applicant quality, suggesting these changes are not driven by changes to the composition of the applicant pool. (See Appendix Section~\ref{sapp:essay_quality} and Fig.~\ref{fig:quality_ts_controlled}.)

\subsection{AI use is penalized}

The widespread availability of AI meaningfully improved the quality of submitted essays, raising the possibility that applicants who use it gain an advantage over those who do not. To investigate this possibility, we next estimate the impact of submitting AI-written essays on admissions decisions.

In the 2023 and 2024 admissions cycles, applicants with at least one essay flagged as AI-written were admitted substantially less often than those without any such essays.\footnote{%
    In the 2025 admissions cycle, after seeing our results, the program we analyze overhauled its admissions process. Given the potential influence of our results on the decision-making process, we do not examine outcomes for that most recent cycle.
}
But applicants who use AI may differ from those who do not in a variety of ways, complicating a causal interpretation of the different admissions rates. By leveraging full application files reviewed by admissions officers, we can account for a wide set of potential confounders, though we cannot completely eliminate the possibility of omitted-variable bias. To adjust our estimates for observable differences between applicants, we apply double machine learning (DML)~\citep{chernozhukov_doubledebiased_2018}, a popular method for estimating causal effects with flexible models that incorporate a large number of covariates---around 400 in our case, after incorporating missingness indicators and other structured applicant information, like age and standardized test scores.

The results are shown in Table~\ref{tab:admissions}. Each column reports the estimated percentage-point change in admission probability for each AI-flagged essay an applicant submitted. The first column reports simple OLS estimates, without adjusting for any applicant characteristics. Our main DML estimates are given in the second and third columns. In the first row of the second column, we estimate a 1.5 p.p.\ decrease in admission probability for each submitted essay that was AI-written, after controlling for our rich set of covariates, but excluding measures of essay quality because of potential post-treatment bias. The rows below show the results from two models that separately estimate the effect for each of the two cycles. The estimated effects are negative in both cycles but more precisely estimated in 2024, likely because AI use was less common in 2023; consequently, the year-specific estimate is not statistically significant in 2023. Combining both cycles, applicants who used AI submitted, on average, 2.32 AI-written essays, corresponding to an aggregate 3.48 p.p.\ decrease in admission probability.

In the third column of Table~\ref{tab:admissions}, we estimate the effect of AI use on admissions decisions after additionally adjusting for the dimensions of essay quality that we constructed above---in addition to the myriad application features we previously adjusted for. This model specification aims to isolate the penalty for AI use itself, accounting for substantive differences in the AI-written and non-AI-written essays, though it may be subject to post-treatment bias. In this case, the estimated effect increases to a penalty of 2.6 p.p.\ for each AI-written essay, or about 6.0 p.p.\ for the typical applicant who used AI. That increase is expected, since AI-written essays are, on average, stronger, and stronger essays themselves improve an applicant's odds of admission. As a result, not accounting for essay strength masks some of the penalty from AI use.

\begin{table}[t]
  \centering
  \normalsize
  \setlength{\tabcolsep}{2.5pt}
  \renewcommand{\arraystretch}{1.08}
  \caption{AI Usage and Admission}
  \label{tab:admissions}
  \begin{tabular*}{\columnwidth}{@{\extracolsep{\fill}}lccc@{}}
    \hline\hline
     & \multicolumn{1}{c}{OLS} & \multicolumn{2}{c}{DML} \\
    \cline{2-4}
     & \multicolumn{1}{c}{(1)} & \multicolumn{1}{c}{(2)} & \multicolumn{1}{c}{(3)} \\
    \hline
    \multirow{2}{*}{AI Count} & -0.077*** & -0.015** & -0.026*** \\
     & (0.007) & (0.006) & (0.006) \\
    \hline
    \multirow{2}{*}{AI Count (2023)} & -0.081*** & -0.015 & -0.021 \\
     & (0.013) & (0.012) & (0.013) \\
    \multirow{2}{*}{AI Count (2024)} & -0.076*** & -0.013* & -0.021** \\
     & (0.008) & (0.006) & (0.007) \\
    \hline
    Applicant controls &  & $\checkmark$ & $\checkmark$ \\
    Essay quality &  &  & $\checkmark$ \\
    \hline\hline
  \end{tabular*}
  \par\medskip
  \parbox{\columnwidth}{%
    \normalfont\emph{Note: Standard errors are reported in parentheses. The first row reports estimates from models without year interactions; the final two rows report estimates from separate year-specific models. Column (1) reports unadjusted OLS estimates from linear probability models. Columns (2) and (3) report DML estimates using random forest nuisance functions and 5-fold cross-fitting. Across 20 independently seeded fits, coefficients and standard errors are aggregated using DoubleML's median-based repeated-splitting procedure~\citep{chernozhukov_doubledebiased_2018}. \({}^{***}\)~\(p < 0.001\), \({}^{**}\)~\(p < 0.01\), \({}^{*}\)~\(p < 0.05\)}
  }
\end{table}

The apparent AI penalty might be expected given that AI use was explicitly prohibited by the program we study. But the admissions office we study did not systematically attempt to detect, adjudicate, or penalize violations of the policy. In particular, submitted essays were not screened with any AI detection tools. Readers might have informally downgraded applicants who used AI, though doing so would require readers to identify AI-written essays.

To understand how well admissions officers can identify AI use in application essays, during the essay quality calibration procedure described above, we additionally asked admissions staff to assess the likelihood that essays were written by a human or by AI, on a six-point scale from ``very likely human'' to ``very likely AI''. (Further details are provided in Appendix Section~\ref{sapp:experiment}.) Admissions staff achieved an AUC of \(0.70\) \((95\% \text{ CI: } [0.65, 0.75])\) on this detection task, meaningfully above chance though far below the accuracy achieved by commercial detectors.

\begin{figure}[t]
\centering
\includegraphics{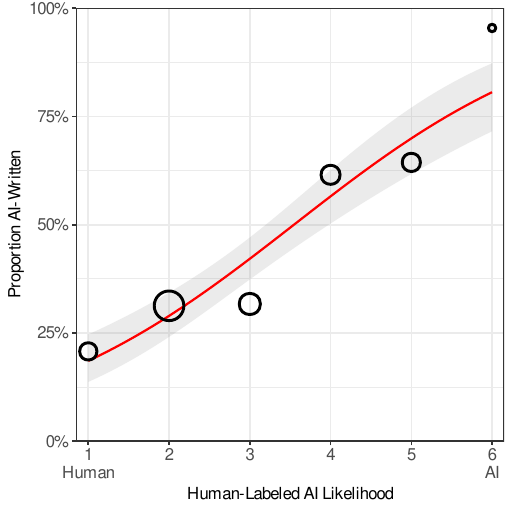}
\caption{%
    \emph{Proportion of essays classified as AI-generated, by admissions officers' perceived likelihood of AI authorship. Point size reflects number of essays in each response category. The red line is a smoothed fit with 95\% confidence bands. The AI-likelihood scale ranges from 1 (``very likely human'') to 6 (``very likely AI''), with intermediate values reflecting increasing likelihood of AI authorship.}
}
\label{fig:human_detection}
\end{figure}

Figure~\ref{fig:human_detection} provides further detail, plotting the relationship between admissions officers' perceived likelihood of AI authorship and the essay's machine-inferred provenance. We find that the human judgments are directionally informative but imprecise. For example, among essays rated on the human side of the scale (1--3), 70.4\% were classified as human-generated by GPTZero, and among those rated on the AI side, 67.8\% were classified as AI-generated by GPTZero. It thus seems that admissions officers, while imperfect, are able to identify AI writing reasonably well.

\begin{figure}[t]
\centering
\includegraphics{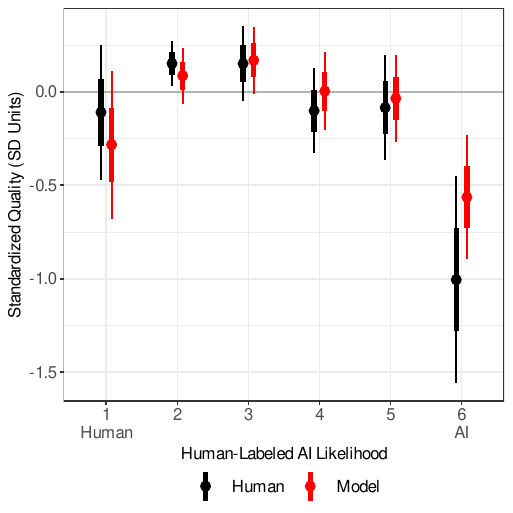}
\caption{%
    \emph{Normalized human and human-calibrated model essay quality scores as a function of admissions officer perception of AI usage. The AI-likelihood scale ranges from 1 (``very likely human'') to 6 (``very likely AI''), with intermediate values reflecting increasing likelihood of AI authorship.}
}
\label{fig:human_and_ai_quality}
\end{figure}

\begin{table}[t]
  \centering
  \normalsize
  \setlength{\tabcolsep}{2.5pt}
  \renewcommand{\arraystretch}{1.08}
  \caption{AI-Likeness and Human Quality Ratings}
  \label{tab:quality_regression}
  \begin{tabular*}{\columnwidth}{@{\extracolsep{\fill}}lccc@{}}
    \hline\hline
     & \multicolumn{3}{c}{Human Quality Rating} \\
    \cline{2-4}
     & \multicolumn{1}{c}{(1)} & \multicolumn{1}{c}{(2)} & \multicolumn{1}{c}{(3)} \\
    \hline
    \multirow{2}{*}{AI Likelihood} & -0.131* & -0.119* & -0.109* \\
     & (0.059) & (0.047) & (0.047) \\
    \multirow{2}{*}{LLM-Based Quality} &  & 1.029*** & 0.995*** \\
     &  & (0.090) & (0.093) \\
    \multirow{2}{*}{AI-Written} &  & 0.019 & -0.033 \\
     &  & (0.101) & (0.102) \\
    \hline
    Applicant controls &  &  & $\checkmark$ \\
    Reviewer fixed effects & $\checkmark$ & $\checkmark$ & $\checkmark$ \\
    \hline\hline
  \end{tabular*}
  \par\medskip
  \parbox{\columnwidth}{%
    \normalfont\emph{Note: Standard errors are reported in parentheses and clustered by applicant. Applicant controls in column (3) include undergraduate GPA, quantitative and verbal GRE scores, residency status, gender, age, and an indicator for whether the applicant submitted a TOEFL score. \({}^{***}\)~\(p < 0.001\), \({}^{**}\)~\(p < 0.01\), \({}^{*}\)~\(p < 0.05\)}
  }
\end{table}

Finally, we consider whether admissions officers assign lower scores to essays they perceive to be AI-generated, a pattern that would be consistent with past work suggesting that AI-writing is penalized in both admissions and more broadly~\citep{digitaldividelee2026,raj_artificial_2026}. The black points in Figure~\ref{fig:human_and_ai_quality} show the average rating assigned to essays as a function of perceived likelihood that they were AI-generated. Essays deemed by admissions staff to be ``very likely'' AI-generated receive markedly lower scores. These essays were also rated lower by our automated evaluation procedure, shown by the red points---though not as low as by admissions staff. However, it is possible that perception of AI correlates with substantive dimensions of essay quality, and those aspects, not perceived AI use, drive the lower ratings. In Table~\ref{tab:quality_regression}, we estimate staff ratings after adjusting for a variety of covariates, including our machine-generated measures of essay quality, whether the essay was in reality AI-generated, and basic demographic and academic information about the applicant. After this adjustment, we still find that staff ratings are negatively correlated with perceived AI use, providing further evidence that AI use itself was penalized by admissions staff. In sum, the lower observed acceptance rate for applicants who use AI is consistent with admission officers detecting and penalizing AI writing. For more detailed discussion of our experiment and reports of individual rater performance, see Appendix Section~\ref{sapp:experiment} and Figs.~\ref{fig:exp_ai_lik_dist} and~\ref{fig:exp_quality_dist}.

\section{Discussion}
Drawing on nearly 7{,}500 applications across six admissions cycles, we document a dramatic shift in applicant behavior. By 2025---just three years after the introduction of ChatGPT---a majority of applicants submitted at least one essay written primarily by AI, despite explicit prohibitions on its use. Among international applicants, the rate approached 70\%. We further find that AI use improved the overall quality of submitted essays, particularly the more mechanical aspects of writing. Yet admissions staff could often recognize AI-written essays and appear to have penalized them.

Our findings expose a central tension for university admissions. AI can improve the quality of submitted essays while weakening the connection between those essays and applicants' unaided writing ability---and potentially subjecting applicants to idiosyncratic penalties for using AI. Even before AI, application essays did not necessarily reflect an applicant's unaided effort, as students frequently turned to mentors, teachers, and others for help. But AI assistants have likely widened the gap between an applicant's individual writing ability and the quality of the final submitted product. At the same time, by reducing the burden of writing, AI may help applicants communicate their ideas more clearly, even as uncritical use of these tools may also distort or displace those ideas. Ultimately, universities must clarify what they seek to learn from application essays and reconsider whether the purposes those essays have historically served remain appropriate in a world where AI-assisted writing is becoming standard professional practice.

These results and their implications should be considered in light of several limitations. First, our analysis is limited to applications to a single graduate program in public policy, and the patterns we document may differ, potentially substantially, across contexts. For example, in settings where AI use is not expressly prohibited, its prevalence may be even higher and its consequences less severe. Second, our estimates of AI use rely on commercial detectors calibrated to limit false positives, and are thus likely conservative. Moreover, to facilitate interpretation, we consider as ``AI generated'' only essays that were likely primarily written by AI, excluding less extensive uses of AI that were also prohibited. Third, our automated measure of essay quality is an imperfect proxy for expert judgment. Although it generally tracks ratings by admissions staff, it may systematically miss dimensions of quality that matter in admissions. Finally, our estimate of the apparent penalty associated with AI use may be affected by unmeasured confounding. Our double machine learning approach adjusts for hundreds of covariates derived from submitted application materials, but we cannot rule out omitted factors associated with both AI use and admissions decisions.

Looking ahead, our findings highlight the need for institutions to move toward deliberate, transparent policies around AI use. Many institutions have yet to establish policies governing AI use in admissions, and even where formal restrictions exist, the boundaries of permissible AI assistance may be unclear and enforcement inconsistent. Our empirical results, on their own, do not determine what role, if any, AI-assisted writing should play in a well-designed admissions process. Rather, by providing systematic evidence on the prevalence and consequences of AI use, we hope to help institutions make more informed decisions about what they seek to learn from applicants' written materials and how those materials should be evaluated in an increasingly AI-mediated world.

\bibliographystyle{unsrtnat-nourl}
\bibliography{refs}

\clearpage
\appendix
\setcounter{figure}{0}
\renewcommand{\thefigure}{A\arabic{figure}}
\renewcommand{\theHfigure}{appendix.figure.\arabic{figure}}

\setcounter{table}{0}
\renewcommand{\thetable}{A\arabic{table}}
\renewcommand{\theHtable}{appendix.table.\arabic{table}}


\section{Supplementary notes on dataset curation}
\label{sapp:notes}
The data collection and featurization pipeline used for our analysis is described in greater detail in concurrent work~\citep{gaebler_test_scores, isley2026mitigatinglabelbiasinterpretable}. Here, we summarize the key elements relevant for this work, and provide further information about unique data used for this paper. In particular, Section~\ref{sapp:data} details the construction of our university application dataset, the covariates we use in our predictive models, and the extraction and featurization pipeline used to structure applicants' transcripts and other admissions materials. Section~\ref{sapp:ai_detectors} discusses how we created our AI usage data. 

\subsection{Admissions data}
\label{sapp:data}
Our data include applications to a two-year public policy master's program at a competitive American professional school submitted between 2020 and 2025, the most recent application cycle, totaling 7{,}462 applications. Each application contains a variety of information about an applicant, including biographical information and applicant metadata, admissions essays, previous or concurrent academic transcripts, resumes, letters of recommendation, and test scores. This set of materials comprises all of the information available to admissions officers at the time of admissions decision-making. 

In collaboration with admissions officers at the university we partner with, we translate this structured and unstructured data into a rich dataset of 427 covariates. Thirty of these features come from pre-structured application data. 216 come from our extraction pipeline described below. We use reported school information to join in another 181 features with information on an applicant's prior institution(s). Before fitting any models, we impute missing values, add indicators for missingness when appropriate, and remove non-varying and collinear columns. Further details are presented in Section~\ref{sapp:admission_penalties}.

\subsubsection{Structured data}

\paragraph{Biographical information and applicant metadata} In filling out their application, applicants report basic information about themselves. This includes concurrent applications to other programs at our partner institution, self-reported age, gender, race, military service, and other demographics. This information is captured in a structured format at the time of application submission, and presented to admissions officers at decision time.\footnote{Following the 2023 Supreme Court decision in \emph{SFFA v.\ Harvard}, legally protected characteristics such as race generally cannot be lawfully considered in admissions in the U.S.~\citep{2023students}. Following this decision, these data have since been unavailable to admissions officers. However, the data are still collected for administrative use.} Admissions officers are also presented with information regarding previous admissions cycles in which the applicant has applied, and whether they were admitted and enrolled.

\paragraph{Test scores}  As part of their application, students are required to submit GRE or GMAT scores.\footnote{In 2020 and 2021, our partner institution suspended testing requirements because of the COVID-19 pandemic.} Additionally, applicants who are not native English speakers and did not attend an English-language undergraduate institution are required to submit IELTS, TOEFL, or Cambridge English scores. Among the 1{,}580 applicants who report language test scores, roughly 63\% submit TOEFL scores and 37\% submit IELTS scores. No students report Cambridge English scores. 

Between the GRE and GMAT, applicants primarily submit GRE scores, from which we include quantitative and verbal section scores in our predictive models. To include in our models the 6.4\% of applicants who submit GMAT scores instead, we convert GMAT quantitative and verbal scores to equivalent percentile scores for the GRE to place GMAT and GRE scores on a common scale~\citep{ets2014gre_concordance}. To compensate for this conversion, we additionally include an indicator for whether an applicant's ``GRE'' scores instead represent a converted GMAT score. 

\paragraph{University rankings} As part of their application materials, applicants are required to list each university at which they previously enrolled, in addition to submitting an official transcript from each school. To supplement our covariate set with metadata on the quality of an applicant's prior education, we leverage the 2025 QS World University Rankings~\citep{qs2025rankings} and IPEDS survey data~\citep{ipeds2024}. The inclusion of these datasets allows us to incorporate university rankings, size, level of research activity, admission rates, graduation rates, financial information, demographic information, average test scores, and employment outcomes as covariates in our models. In total, we consider 44 features from IPEDS and 46 from QS data. In our models, we include this data for \textit{both} their previous undergraduate and graduate degrees separately. Additionally, from the institutional information available in the IPEDS survey data, we predict QS rankings for schools with otherwise missing rankings. We split these rankings (real or predicted) into quintiles, and include that quintile as an additional feature, granting a total of 181 features from IPEDS and QS data. 

\subsubsection{Unstructured data}

\paragraph{Transcripts}

Transcripts provide a rich source of information about an applicant's academic performance at previous institutions attended. However, the variety of formats and grading conventions makes them difficult to process at the scale required for predictive modeling. To incorporate this information into our models, we created the following process for extracting features from this valuable unstructured data. 

The pipeline for processing transcripts consists of four steps: 
\begin{enumerate}
    \item First, we preprocess the transcript files to enhance the quality of our subsequent extraction. In particular, we use the multimodal capabilities of OpenAI's \texttt{gpt-5-nano}~\citep{openai2025gpt5} to identify the rotation of each page in the transcript. We un-rotate any rotated pages so that the text reads horizontally. We additionally increase the text contrast on each page and remove any watermarks. 
    \item Once preprocessed, each transcript is transcribed to markdown via Microsoft Document Intelligence~\citep{microsoft2024documentintelligence}. 
    \item Then, we prompt \texttt{gpt-5-mini} to extract course-level information (e.g., term, title, course codes, credits, grade, instructor name, and level) and further term- and degree-specific information (e.g., official institution name and address, matriculation and graduation dates, and majors and minors) in a loosely structured format, primarily to remove extraneous information and reduce errors downstream. 
    \item Finally, this loosely structured representation of an applicant's transcript is converted into structured output in accordance with a fixed JSON schema, again by \texttt{gpt-5-mini}.
\end{enumerate}

After processing the raw transcript files, we then featurize this newly structured data into a set of 155 measures of previous academic performance. These include overall GPAs, subject area GPAs (e.g., mathematics and statistics, economics), course levels (e.g., upper- and lower-undergraduate), GPA trends over time, performance in specific courses (e.g., introductory micro- and macroeconomics), indicators for majors, minors, and honors, and other similar measures. For applicants with multiple transcripts, we calculate credit-weighted GPAs across institutions. 

To ensure a constant 4.0 scale across all applicants in our sample, approximately half of whom come from international institutions, we prompt \texttt{gpt-5-mini} in Step 4 to convert the GPA to a 4.0 scale, if a conversion key was extracted from the raw transcript. If not, we include in our prompt all available conversion keys from the Scholaro database~\citep{scholaro2025} for the relevant country (the same resource our partner admissions office uses to convert GPAs when no scale is provided). 

\paragraph{Letters of recommendation} Letters of recommendation are submitted in a more standardized format than transcripts. The applicant must submit a total of three letters of recommendation from prior professors, professional supervisors, or other individuals who can speak to the applicant's readiness for the program. Together with our partner admissions office, we developed two rubrics for evaluating the written text submitted in the letter. The first focuses on the relevance of a recommender's relationship to the applicant, the strength of their recommendation, the recommender's profession, seniority, and employing organization, their apparent depth of knowledge about the candidate and the length of their relationship, as well as the letter writer's account of the applicant's academic preparedness, professional leadership, commitment to public service, and curiosity. The other rubric focuses specifically on whether the letter makes specific reference to applied quantitative skills or other evidence of quantitative preparation for the program.

In addition to the written letter, recommenders also submit a cover sheet, providing information about themselves, their relationship with the applicant, the number of other letters they are writing, and the individuals against whom the applicant is being compared. Additionally, recommenders answer Likert-style questions about an applicant's academic preparation, ability to express themselves, work ethic, personality traits, and other dimensions of their preparedness for graduate school. 

As with transcripts, we convert an applicant's three letter/cover sheet pairs into markdown using Microsoft Document Intelligence~\citep{microsoft2024documentintelligence}. In three separate prompts, we then prompt \texttt{gpt-5-nano} to extract the structured applicant ratings from the cover sheet and evaluate the letter according to our two rubrics. These three extractions grant a total of 39 measures related to an applicant's letters of recommendation. 

\paragraph{Resumes} Once again, we leverage Microsoft Document Intelligence~\citep{microsoft2024documentintelligence} to convert an applicant's submitted resume PDF into markdown text. We then prompt \texttt{gpt-5-nano} to structure the applicant's work history. To do so, for each position listed in their resume, we extract the applicant's title and tenure, employment status (intern, part-time, full-time) and seniority, sector (for-profit, non-profit, government), size and prominence of their employer, leadership, public service, or quantitative responsibilities, and other key information. We also prompt the model to classify the applicant's career trajectory according to the promotions, demotions, and lateral career moves contained in their resume. From each applicant's structured work history, we calculate 8 summary measures of program-relevant work experience, such as their number of years of work in the public sector. 

\paragraph{Essays} 
\label{sapp:essay_quality_data}
Finally, as part of their application, applicants are required to submit at least five essays of around 250 to 500 words, covering topics like their interest in the specific program, the unique perspective they bring to the school, and their interest in public service. This core set of essay topics remains constant throughout our six-year sample, barring one essay. That essay's prompt changes significantly in 2023, though the core question remains qualitatively similar.\footnote{We reproduce our analysis without this essay in Section~\ref{sapp:robustness}, and find qualitatively similar results.}

To evaluate the quality of these essays in a way that matches that of an admissions officer, we develop a rubric that mirrors the rubric used in the admissions process to evaluate seven aspects of the essays: their grammatical correctness, stylistic quality, clarity and organization, adherence to the prompt and formal requirements, voice and originality, public service merit, and evidence of leadership potential. To do so, after converting to markdown with Microsoft Document Intelligence~\citep{microsoft2024documentintelligence}, we again prompt \texttt{gpt-5-mini} to grade each essay according to our rubric, leveraging concrete evaluation criteria for each dimension. For instance, grammatical quality is scored according to the number and severity of errors. We rate evidence of leadership potential in accordance with whether the applicant mentioned a specific future leadership role, and the number of details they provided about that position and their plans to attain it.

Additionally, admissions officers report placing particular emphasis on the contents of one of the five essays. To account for this reported heightened attention to this essay, we include the LLM-assessed quality scores for that essay as an additional seven features, granting 14 total essay quality features.

\subsubsection{Featurization}
After compiling this expansive list of applicant covariates, we adopt standard machine learning approaches to handle data missingness. In particular, we impute any missing numerical covariates with their median, and impute any missing logical covariates with \texttt{FALSE}. To allow the model to learn if missingness itself is important, we additionally include a missingness indicator for features where missingness may be informative. In total, from our featurization pipeline, we are left with 427 complete covariates and 420 missingness indicators.

\subsubsection{Outcomes}
Our two primary outcomes of concern are essay quality and admission probability. In Section~\ref{sapp:essay_quality_data} and in Section~\ref{sapp:essay_quality}, we detail our process for generating essay quality ratings, which, as we show in the main text (Figure~\ref{fig:human_model_calib}), align well with human ratings of essay quality. Therefore here we only detail our admissions-related outcome. 

In addition to each applicant's application materials, we have access to their admissions decisions. We use a binary indicator of whether or not the applicant was admitted to the program as our main outcome of interest. Importantly, though we have access to application materials from the most recent round (submitted in 2025), the admissions office we partner with substantially changed its admissions process in light of our preliminary results, and thus we do not analyze outcomes for that cycle. 

\subsection{AI detection}
\label{sapp:ai_detectors}
Because we cannot observe true AI authorship in the application data, our empirical approach depends on the accuracy of AI detection tools, as we use their detection to serve as a proxy for true AI usage. To minimize measurement error in our estimates of potential benefits or penalties associated with AI usage, we explore two detectors recently shown to have impressively low false positive and false negative rates: GPTZero and Pangram~\citep{jabarian_artificial_2025, adam2026gptzerorobustdetectionllmgenerated, emi_technical_2024}. In our main analysis, we consider GPTZero's model \texttt{3.9}. In our robustness checks, we use Pangram's version \texttt{3.1}. 

\subsubsection{Additional text cleaning} Before running an essay through any of our detectors, we complete an additional layer of text cleaning to remove any markdown artifacts from Microsoft Document Intelligence~\citep{microsoft2024documentintelligence} and to strip any headers or essay prompts from the essay text, ensuring the AI detection classifiers only run on the applicant's submitted writing. To do so, we prompt \texttt{gpt-5-nano}~\cite{openai2025gpt5} to reproduce each essay without any markdown tags, student metadata included in the document's header, or prompt text. To ensure the model did not alter the text of the applicant's essay, we validate the cleaning by computing character-level diffs between the original and cleaned versions. For any essay with a non-zero diff, we prompt \texttt{gpt-5-nano} to flag any instances where the cleaning introduced errors, altered meaning, or changed the original text. We manually review all flagged cases $(n \approx 1200)$, and confirm changes made were strictly subtractive of non-essay text, manually reverting LLM-made changes for a small number of essays. In addition to validating that the LLM-cleaner did not alter the essay text, we further confirm the resulting labels do not change significantly. Following cleaning, labels change for 1{,}015 (2.7\%) essays, with a large majority (865) changing classification from AI to non-AI. 

\subsubsection{Detector error rate validation} As discussed in the main text, GPTZero classifies text into one of three provenances: ``AI,'' ``Mixed,'' or ``Human.'' GPTZero additionally reports the confidence of their classification as ``Low,'' ``Medium,'' or ``High.'' The other detector we consider here, Pangram, uses similar ``AI,'' ``Human,'' and ``Mixed'' labels, but does not provide document level confidence assessments. To determine which, if any, of these provenances achieved an error rate appropriate for our analysis, we first investigated each detector's error rates on our data for a variety of different thresholds signifying ``AI use.'' 

\paragraph{False positive rates} Leveraging our historical data, we confirm the false positive rates of each detector on the essays submitted prior to the release of ChatGPT.\footnote{%
    The application deadline for applications submitted in 2022 was one day after November 30th, so we assume no essays submitted in 2022 were written using LLMs. While large language models existed prior to the commercial release of ChatGPT \citep[e.g. as in][]{brown_language_2020}, we assume their use in applicant essays is plausibly negligible in practice, as models were less accessible to broad audiences at the time.
} 
Tables~\ref{tab:fpr_detection_app} and~\ref{tab:fpr_detection_essay} report the AI detection rates by year of material submission across a variety of classification thresholds, at the applicant and essay level, respectively. At the essay level, both GPTZero and Pangram display impressively low false positive rates for their ``AI'' labels, matching prior analyses of AI detectors~\citep{jabarian_artificial_2025}. For GPTZero, this low false positive rate increases marginally when the required confidence threshold is relaxed, and slightly more significantly when ``mixed'' labels are considered as AI usage. When both ``mixed'' labels are used and confidence levels are relaxed, the essay-level false positive rate increases by an order of magnitude. Pangram also displays a much higher false positive rate when ``mixed'' labels are considered. As applicants submit multiple essays in their application, these false positive rates increase when aggregated to the applicant level, as expected, though the general patterns remain the same.

\paragraph{False negative rates}
\label{sapp:fnr_discussion}
In our context, we cannot meaningfully estimate false negative rates, as we do not have access to ground truth applicant AI usage. To the extent to which AI usage goes undetected, non-AI users in our data will include some true AI users, and therefore our estimates may reflect the effect of detectable AI usage rather than AI usage at large. However, prior work reports low false negative rates on text entirely written by AI~\citep{jabarian_artificial_2025}---precisely the provenance captured by GPTZero and Pangram's ``AI'' labels. This suggests that our labels are unlikely to systematically miss essays that were fully AI-authored, limiting measurement error concerns for the treatment we consider in our main analysis. 

\subsubsection{Detector selection}
\label{sapp:detector_selection}
As our primary estimands are applicant-level, we base our detector selection on applicant-level false positive rates. As shown in Table~\ref{tab:fpr_detection_app}, only GPTZero ``AI'' labels with ``high'' confidence and Pangram ``AI'' labels maintain false positive rates below 1\% (on average) on the pre-ChatGPT data (0.8\% and 0.3\%, respectively). To prevent overstating our prevalence estimates and to mitigate noise in treatment identification, we therefore only consider these two thresholds as indicative of AI authorship in our main analysis. We selected GPTZero for our main text results, though Pangram's ``AI'' labels produce qualitatively similar results. Section~\ref{sapp:robustness} provides more commentary on the robustness of our results to different threshold choices and AI detection software. 

\subsubsection{Usage by demographics} 
\begin{figure}
\centering
\includegraphics{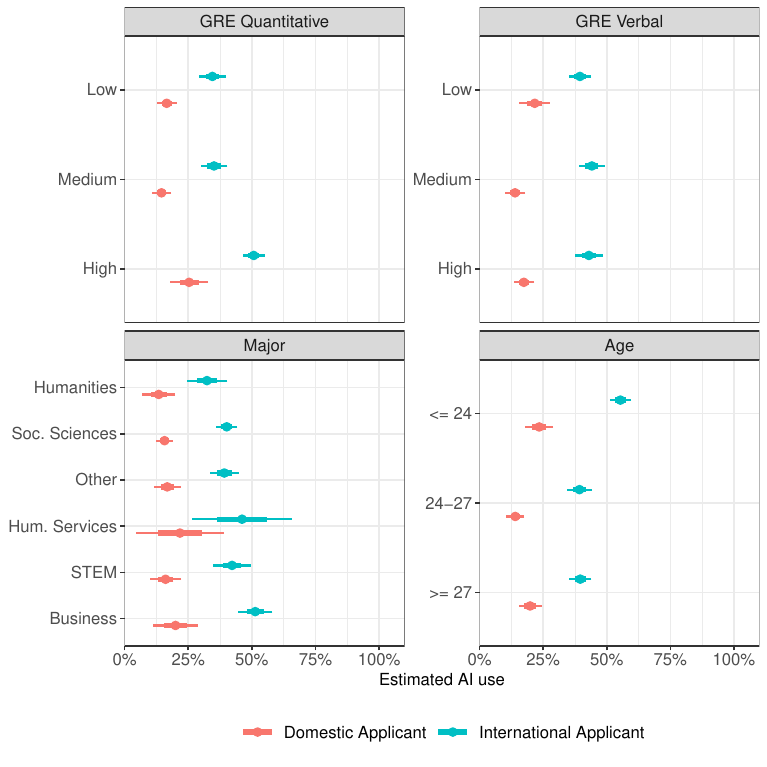}
\caption{GPTZero estimated applicant-level AI usage by demographic
characteristics, disaggregated by nationality. The ``Low,'' ``Medium,'' and
``High'' bins in the GRE plots contain the lower, middle, and upper tertiles
of GRE scores, respectively. Age bins each contain roughly one third of the
applicant pool.}
\label{fig:usage_by_demo}
\end{figure}

Figure~\ref{fig:usage_by_demo} reports AI usage rates across demographic subgroups. Most strikingly, we find a large and persistent gap in usage between domestic and international applicants. International applicants show substantially higher AI usage across every subgroup we examine. This gap is most pronounced for applicants below the age of 24. AI usage also appears to be higher among international applicants who score highly on the quantitative section of the GRE, though not among domestic applicants. AI usage is flat for both groups across verbal scores, suggesting AI usage is not concentrated among weaker applicants.

\section{Experimental details}
\label{sapp:experiment}
To assess how well admissions officers can detect AI-written content and to better understand how they determine an essay's quality, we ran an experiment on the admissions officers at our partner university. (Our study, protocol number IRB24-0242, was granted IRB approval by the Harvard University-Area Committee on the Use of Human Subjects. Consent was obtained in person, and data were analyzed anonymously.) From the admissions office, we recruited five staff members who were going to review applications in the upcoming admissions cycle (2025--2026 cycle). We selected 100 essays (50 human, 50 AI) submitted in either 2023 or 2024, all in response to the same prompt. To ensure AI and human essays came from applicants with similar observable profiles, we used \texttt{MatchIt}~\citep{matchit} to construct a 1:1 nearest-neighbor propensity score matching on applicant characteristics. We then randomly sampled 50 matched pairs from the resulting set, granting 50 human- and 50 AI-written essays from applicants with similar observable characteristics.

Each officer was tasked to review all 100 essays sequentially, in a random order, with the option to pause and resume their reviews at any time. After reading each essay, participants were asked to rank, on a six-point Likert scale, both how likely they thought the essay was AI-generated and how they would rate its quality in comparison to essays they had reviewed as part of the admissions process. The options on the AI-likelihood scale included: ``Very likely human,'' ``Moderately likely human,'' ``Somewhat likely human,'' ``Somewhat likely AI,''  ``Moderately likely AI,'' and ``Very likely AI.'' The options on essay quality scale included: ``Much weaker,'' ``Moderately weaker,'' ``Somewhat weaker,'' ``Somewhat stronger,'' ``Moderately stronger,'' and ``Much stronger.''

\begin{figure}
\centering
\includegraphics{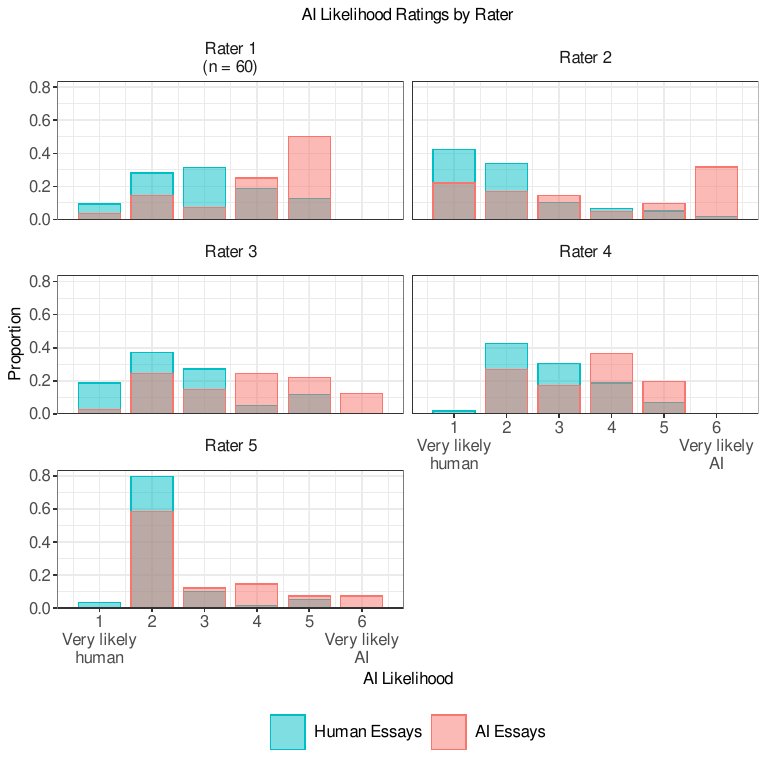}
\caption{Distribution of AI likelihood ratings by rater, disaggregated by essay provenance. Each panel shows one admissions officer's ratings on a six-point scale ranging from 1 (``Very likely human'') to 6 (``Very likely AI''). Bars show the proportion of AI-written and human-written essays receiving each rating. Rater 1 completed 60 of the 100 essays.}
\label{fig:exp_ai_lik_dist}
\end{figure}

\begin{figure}
\centering
\includegraphics{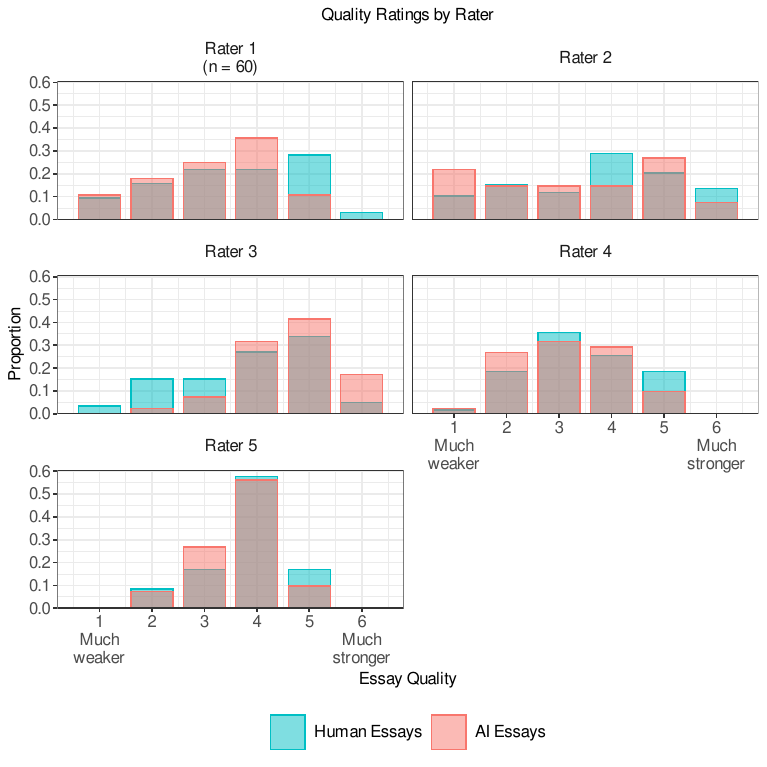}
\caption{Distribution of essay quality ratings by rater, disaggregated by essay provenance. Each panel shows one admissions officer's ratings on a six-point scale ranging from 1 (``Much weaker'') to 6 (``Much stronger'') relative to a typical applicant. Bars show the proportion of AI-written and human-written essays receiving each rating. Rater 1 completed 60 of the 100 essays.}
\label{fig:exp_quality_dist}
\end{figure}

Of the participants, Raters 2--5 rated all 100 essays. Rater 1 only rated 60 (32 human, 28 AI), which grants a total of 460 ratings. Figures~\ref{fig:exp_ai_lik_dist} and~\ref{fig:exp_quality_dist} show the distributions of each rater's AI likelihood ratings and essay quality ratings, respectively. Most raters demonstrate the ability to discriminate between AI and human essays, though imperfectly: Human essays cluster at lower values and AI essays receive higher ratings on average. That said, Rater 5 assigned nearly all essays a rating of 2 on the AI likelihood scale regardless of provenance, suggesting limited ability or willingness to discriminate between AI and human text. In terms of quality, ratings are more uniform across raters and essay types, consistent with quality being a more subjective judgment. 

\section{Essay quality}
\label{sapp:essay_quality}
Here, we provide further detail on our results pertaining to essay quality. This section contains two parts. First, we detail how we construct human-calibrated quality for all essays in our dataset. Then, we discuss our replication of the results of the essay quality time series analysis, controlling for applicant observables. 

\subsection{Human calibrated essay quality}

After completing the experiment detailed in Section~\ref{sapp:experiment}, we fit a linear regression model in R to extract human weights on our different LLM-extracted dimensions of essay quality. As we report in the main text, we observe a harsh quality penalty when a rater believes an essay to be ``very likely AI,'' so we drop those ratings $(n=22)$ to remove that bias from our estimates. We then regress each rater's quality rating on our seven LLM-assessed dimensions of essay quality, with a rater fixed effect to absorb individual differences in the raters. Table~\ref{tab:quality_weights} displays the coefficients on each quality dimension as estimated by this model. We use these coefficients as weights for computing our human-calibrated composite quality score, used throughout the main text. As Figure~\ref{fig:human_model_calib} shows, our human-calibrated composite reflects human preferences well. 

\subsection{The effect of AI use on essay quality} 
\begin{figure}
\centering
\includegraphics[]{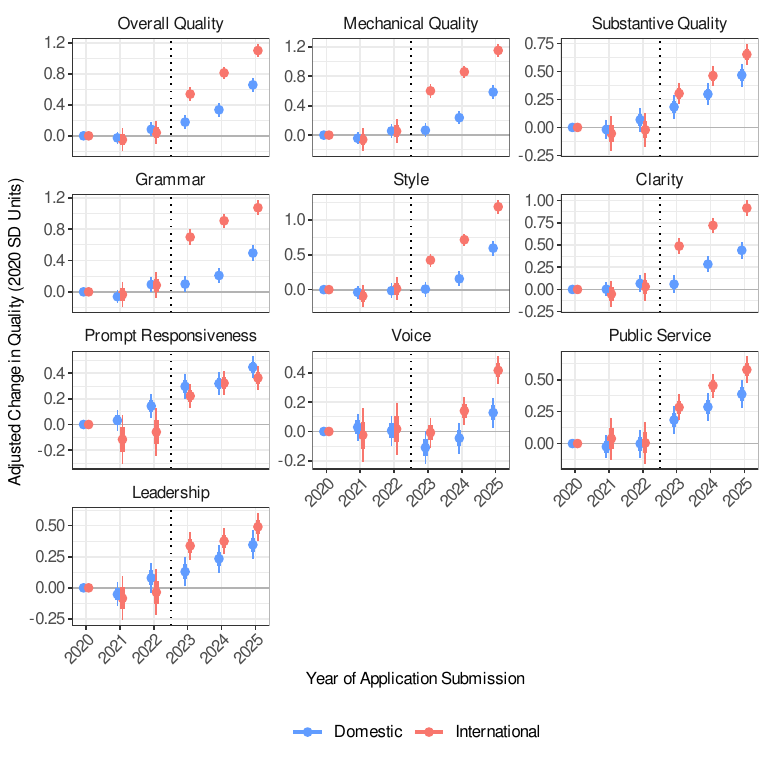}
\caption{Change in LLM-assessed essay quality over time, disaggregated by residency status, conditional on applicant covariates. Points show OLS estimates of round coefficients relative to applicants in 2020, normalized to 2020 SD units. Error bars show $\pm$1 and $\pm$2 SE, clustered at the applicant level. Baseline year (2020) is normalized to zero within group. The dashed line marks the release of ChatGPT in late 2022.}
\label{fig:quality_ts_controlled}
\end{figure}

In the main text, we report a timeseries of essay quality over the six years in our sample, to demonstrate the large improvement in writing quality we observe following the release of ChatGPT. However, it is possible that these gains were instead driven by a shift in the applicant pool. To explore this counterfactual, we produce a similar time series to that of Figure~\ref{fig:quality_over_time} for each dimension of LLM-assessed essay quality while controlling for the rich set of applicant observables described in Section~\ref{sapp:data}, except for essay quality ratings. We estimate separate models for domestic and international applicants. Instead of the group means as reported in Figure~\ref{fig:quality_over_time}, Figure~\ref{fig:quality_ts_controlled} reports the coefficients for each year in those models, providing estimates of essay quality by year controlling for applicant characteristics. These coefficients are then normalized by the 2020 \textit{within-group} standard deviation to ensure a common scale across years and quality dimensions, and track changes relative to the group's baseline. Models are estimated using \texttt{feols} from the \texttt{fixest} package~\citep{fixest} in R. 

We find the gains in essay quality reported in the main text persist after controlling for applicant observables, suggesting the trends observed in Figure~\ref{fig:quality_over_time} are not driven solely by changes in the composition of the applicant pool. Pre-ChatGPT estimates for mechanical quality (grammar, style and clarity), are statistically indistinguishable from zero, followed by dramatic post-ChatGPT gains---particularly for international applicants, whose mechanical gains are roughly double those of domestic applicants by 2025. Substantive dimensions also trend upward, however some (e.g., leadership, prompt responsiveness, and public service) exhibit trends that begin prior to the release of ChatGPT, making attribution to AI-use less clear. These patterns are most consistent with AI improving the mechanical polish of essays, particularly for international applicants, with less clear evidence of the effects on substantive content.

\section{Admissions penalties}
\label{sapp:admission_penalties}
In this section, we provide further details on our DML estimates of the effect of AI usage on admissions outcomes. We estimate partially linear models using the \texttt{DoubleML} R package~\citep{doubleml_package, chernozhukov_doubledebiased_2018}. Our treatment variable is the number of an applicant's essays classified as AI-written, and the outcome is an indicator for admission. While in our prevalence estimates we define AI users as applicants who submitted at least one AI-written essay, in the real admissions process (without access to AI detection tools) admissions officers' demonstrated imperfect detection ability suggests their detection signal of AI use likely scales in the number of AI-written essays submitted by an applicant. To capture this effect, we model a continuous count treatment. The coefficient from these models thus represents the estimated change in admission probability associated with one additional essay classified as AI-written, conditional on observable characteristics. To confirm our results are not an artifact of our continuous treatment effect, we estimate a binary treatment effect of \textit{any} AI usage in Section~\ref{sapp:robustness}, and find qualitatively similar results. 

We report both pooled and admissions-cycle specific estimates for applications submitted in 2023 and 2024. The pooled model combines these two cohorts, and includes an admissions cycle indicator among other applicant-level controls. The cycle-specific estimates are fit only on that cycle's data (e.g., the coefficient for AI Count (2023) is only estimated on the 2023 applicant pool.) We fit separate nuisance functions per cycle. 

The estimates presented in columns (2) and (3) of Table~\ref{tab:admissions} in the main text arise from two highly similar specifications. Column (2) controls for the large array of covariates described in Section~\ref{sapp:data}, but excludes essay quality measures. Because AI usage may affect essay quality, conditioning on these measures could introduce post-treatment bias. Column (3) includes these 14 controls as a descriptive check. While the post-treatment bias complicates the causal interpretation of these estimates, that the penalty grows larger after adjusting for essay quality suggests the admissions penalty is not explained by differences in essay quality between AI and non-AI users. 

Before fitting any models, we first eliminate any columns that are single valued, eliminate any remaining collinear columns by fitting a linear regression on the full covariate set, dropping any features whose coefficients could not be estimated, and finally remove any columns that do not exhibit variation in both 2023 and 2024, separately. After these exclusions, we retain a common set of 411 covariates that exhibit variation across both admissions cohorts, 70 of which are missingness indicators. Our primary specification excludes the 14 essay-level quality measures, and thus includes 397 features. In the pooled models, we add an application cycle term that increases both counts to 412 and 398, respectively.

We estimate the treatment nuisance function and the outcome nuisance function using random forests via \texttt{mlr3} with \texttt{ranger} backend~\citep{ranger_package, mlr3}. We tune the two nuisance functions separately for each estimation sample and covariate specification using 5-fold cross validation. To select hyperparameters, we search over 30 random combinations of \texttt{mtry}, minimum node size, maximum tree depth, and sampling fraction using 200 tree forests. We then fit the final nuisance models using the selected hyperparameters and 500 trees, with 5-fold cross-fitting. 

\paragraph{Cross-seed aggregation.} To account for simulation variability introduced by the random cross-fitting fold assignment, we repeat each DML specification across 20 random seeds and aggregate the results using the repeated splitting procedure implemented by the DoubleML package~\citep{doubleml_package}. All DML estimates, both in the main text and those provided in Section~\ref{sapp:robustness} reflect this repeated-splitting aggregation procedure.

\paragraph{Residual variation checks.} Identification of the partially linear model requires the AI essay count to retain variation after conditioning on our included sets of covariates. Table~\ref{tab:residuals} reports the residual variance and cross-fitted $R^2$ for each treatment nuisance model. For the specifications presented in the main text, residual variances range from 0.80 to 2.04, and cross-fitted $R^2 \leq 0.32$, indicating ample residual treatment variation after conditioning on our covariate set. Across all of our continuous-treatment robustness checks, we similarly find sufficient residual variation. For our binary treatment effect robustness check, we report overlap plots in Figures~\ref{fig:overlap_gz_ai_binary} and~\ref{fig:overlap_pang_binary}.

\section{Robustness}
\label{sapp:robustness}
\begin{figure}
\centering
\includegraphics{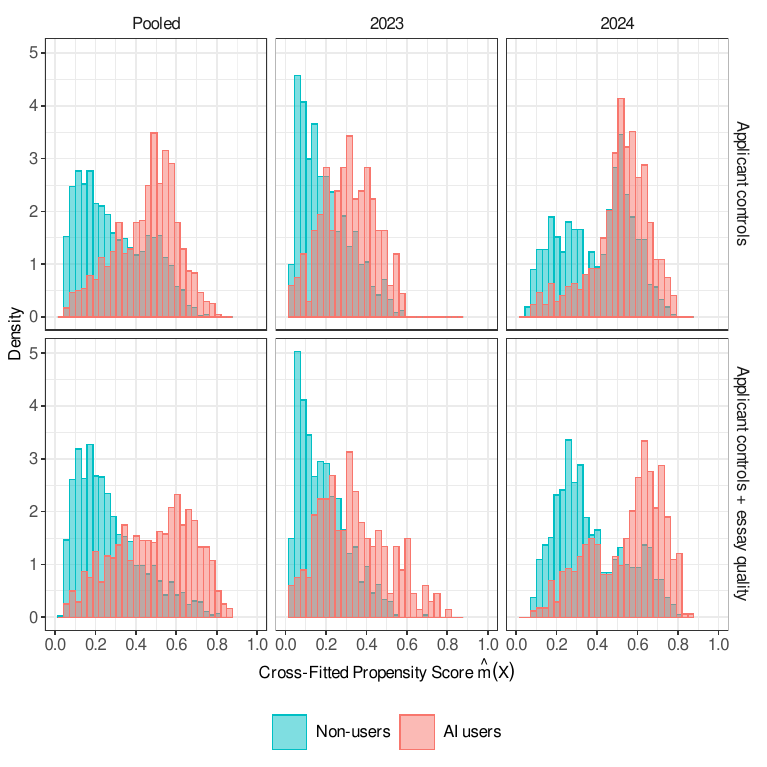}
\caption{Overlap diagnostics for the GPTZero binary treatment specification. Each panel shows the distribution of $\hat P( D =1 | X)$ from the DML treatment nuisance model (random forest), for AI-using and non-AI-using applicants. Columns correspond to application cohort (2023, 2024), rows correspond to specification.}
\label{fig:overlap_gz_ai_binary}
\end{figure}

\begin{figure}
\centering
\includegraphics{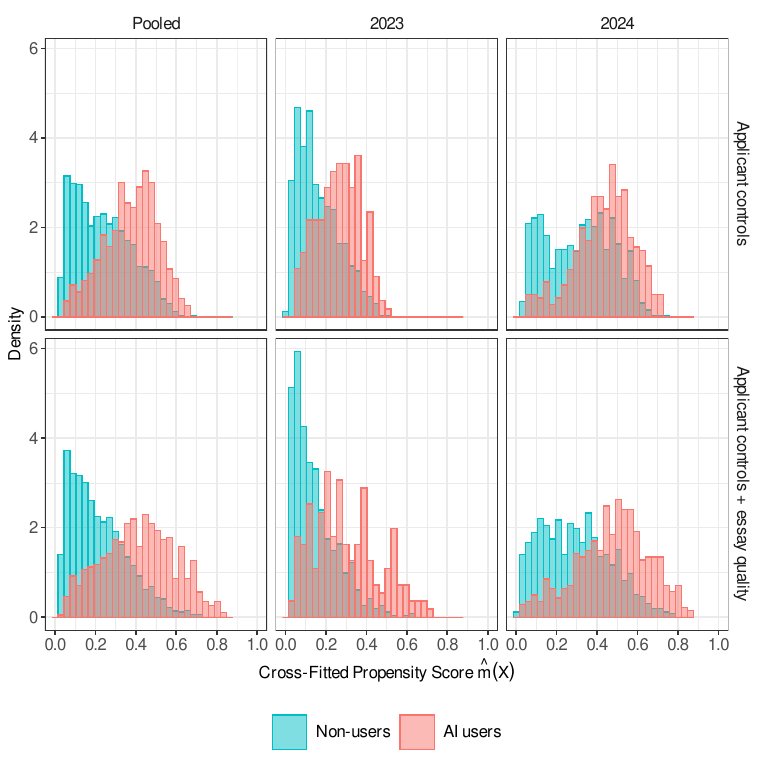}
\caption{Overlap diagnostics for the Pangram binary treatment specification. Each panel shows the distribution of $\hat P(D =1 | X)$ from the DML treatment nuisance model (random forest), for AI-using and non-AI-using applicants. Columns correspond to application cohort (2023, 2024), rows correspond to specification.}
\label{fig:overlap_pang_binary}
\end{figure}

To evaluate the robustness of our admissions results, we vary our analysis along four dimensions. Across all of these dimensions of variation, our results persist. 
\begin{enumerate}
     \item \textbf{Detector choice:} We first vary which AI detection software's labels we consider. Table~\ref{tab:admissions_pangram} replicates our main results using Pangram~\citep{emi_technical_2024}. The effects are qualitatively identical to those we find with GPTZero. 
     \item \textbf{Detection threshold:} In addition to investigating robustness for different detection software, we additionally report estimates for looser ``AI generated'' classification thresholds. In particular, columns (1)--(4) in Table~\ref{tab:dml_robustness} report estimates of the AI-use penalty when we include ``mixed'' label classifications or, in the case of GPTZero, lower confidence classifications. Our pooled estimates are stable across all variations, whereas the year-specific estimates are noisier. 
     \item \textbf{Omission of essay with changed prompt:} As mentioned in the main text, one essay prompt changed significantly in 2023. In Table~\ref{tab:dml_robustness}, column (5), we reproduce our main results excluding this essay entirely. We find qualitatively similar estimates, suggesting the change in this essay's prompt is not impacting our estimated effect. 
    \item \textbf{Binary treatment:} Finally, we replace the count treatment with a binary indicator for whether an applicant submitted any AI-written essay. As seen in Table~\ref{tab:dml_binary}, we find a penalty on any AI usage that is consistent with our main result. We provide support of the overlap assumption for binary treatment in Figures~\ref{fig:overlap_gz_ai_binary} and~\ref{fig:overlap_pang_binary}. 
\end{enumerate}

\FloatBarrier


\begin{landscape}
\begin{table}[p]
  \centering
  \footnotesize
  \setlength{\tabcolsep}{1.2pt}
  \renewcommand{\arraystretch}{1.08}
  \caption{AI Detection Rates by Application Year and Detector (Applicant Level)}
  \label{tab:fpr_detection_app}
  \begin{tabular*}{\columnwidth}{@{\extracolsep{\fill}}lccccccc@{}}
    \hline\hline
    \noalign{\vskip 3pt}
    Year & \makecell[c]{GPTZero \\[2pt] AI Only \\[2pt] High Confidence} & \makecell[c]{GPTZero \\[2pt] AI + Mixed \\[2pt] High Confidence} & \makecell[c]{GPTZero \\[2pt] AI Only \\[2pt] Any Confidence} & \makecell[c]{GPTZero \\[2pt] AI + Mixed \\[2pt] Any Confidence} & \makecell[c]{Pangram \\[2pt] AI Only} & \makecell[c]{Pangram \\[2pt] AI + Mixed} & \makecell[c]{GPTZero \\[2pt] AI Only \\[2pt] Excl. Changed Essay} \\[3pt]
     & \multicolumn{1}{c}{(1)} & \multicolumn{1}{c}{(2)} & \multicolumn{1}{c}{(3)} & \multicolumn{1}{c}{(4)} & \multicolumn{1}{c}{(5)} & \multicolumn{1}{c}{(6)} & \multicolumn{1}{c}{(7)} \\[2pt]
    \hline
    \noalign{\vskip 2pt}
    \multicolumn{8}{l}{\textit{False-positive rates (pre-ChatGPT)}} \\[2pt]
    \textit{2020} & \textit{0.7\% (0.2\%)} & \textit{4.0\% (0.5\%)} & \textit{1.7\% (0.3\%)} & \textit{6.4\% (0.6\%)} & \textit{0.2\% (0.1\%)} & \textit{6.3\% (0.6\%)} & \textit{0.5\% (0.2\%)} \\
    \textit{2021} & \textit{0.5\% (0.2\%)} & \textit{2.9\% (0.5\%)} & \textit{1.3\% (0.3\%)} & \textit{4.7\% (0.6\%)} & \textit{0.4\% (0.2\%)} & \textit{7.2\% (0.7\%)} & \textit{0.5\% (0.2\%)} \\
    \textit{2022} & \textit{1.4\% (0.4\%)} & \textit{4.3\% (0.7\%)} & \textit{1.8\% (0.5\%)} & \textit{6.8\% (0.9\%)} & \textit{0.5\% (0.2\%)} & \textit{11.9\% (1.1\%)} & \textit{0.8\% (0.3\%)} \\
    \hline
    \noalign{\vskip 2pt}
    \multicolumn{8}{l}{\textit{Detection rates (post-ChatGPT)}} \\[2pt]
    2023 & 21.8\% (1.3\%) & 38.2\% (1.5\%) & 25.8\% (1.3\%) & 43.2\% (1.5\%) & 17.4\% (1.2\%) & 47.1\% (1.5\%) & 18.0\% (1.2\%) \\
    2024 & 45.1\% (1.4\%) & 62.6\% (1.3\%) & 47.9\% (1.4\%) & 65.9\% (1.3\%) & 35.3\% (1.3\%) & 64.7\% (1.3\%) & 39.3\% (1.3\%) \\
    2025 & 56.1\% (1.5\%) & 71.5\% (1.3\%) & 60.0\% (1.5\%) & 75.6\% (1.3\%) & 45.9\% (1.5\%) & 63.0\% (1.4\%) & 50.4\% (1.5\%) \\
    \hline\hline
  \end{tabular*}
  \par\medskip
  \parbox{\columnwidth}{%
    \normalfont\emph{Note: Entries report the percentage of applicants classified as AI users, with binomial standard errors in parentheses. An applicant is classified as an AI user if at least one submitted essay satisfies the detector definition shown in the column heading. Italicized rows correspond to pre-ChatGPT application cycles and are interpreted as false-positive rates. Application year refers to the year in which the application was submitted. Column (7) excludes the changed essay from the classification.}
  }
\end{table}

\begin{table}[p]
  \centering
  \footnotesize
  \setlength{\tabcolsep}{1.2pt}
  \renewcommand{\arraystretch}{1.08}
  \caption{AI Detection Rates by Application Year and Detector (Essay Level)}
  \label{tab:fpr_detection_essay}
  \begin{tabular*}{\columnwidth}{@{\extracolsep{\fill}}lccccccc@{}}
    \hline\hline
    \noalign{\vskip 3pt}
    Year & \makecell[c]{GPTZero \\[2pt] AI Only \\[2pt] High Confidence} & \makecell[c]{GPTZero \\[2pt] AI + Mixed \\[2pt] High Confidence} & \makecell[c]{GPTZero \\[2pt] AI Only \\[2pt] Any Confidence} & \makecell[c]{GPTZero \\[2pt] AI + Mixed \\[2pt] Any Confidence} & \makecell[c]{Pangram \\[2pt] AI Only} & \makecell[c]{Pangram \\[2pt] AI + Mixed} & \makecell[c]{GPTZero \\[2pt] AI Only \\[2pt] Excl. Changed Essay} \\[3pt]
     & \multicolumn{1}{c}{(1)} & \multicolumn{1}{c}{(2)} & \multicolumn{1}{c}{(3)} & \multicolumn{1}{c}{(4)} & \multicolumn{1}{c}{(5)} & \multicolumn{1}{c}{(6)} & \multicolumn{1}{c}{(7)} \\[2pt]
    \hline
    \noalign{\vskip 2pt}
    \multicolumn{8}{l}{\textit{False-positive rates (pre-ChatGPT)}} \\[2pt]
    \textit{2020} & \textit{0.1\% (0.0\%)} & \textit{0.8\% (0.1\%)} & \textit{0.3\% (0.1\%)} & \textit{1.4\% (0.1\%)} & \textit{0.0\% (0.0\%)} & \textit{1.4\% (0.1\%)} & \textit{0.1\% (0.0\%)} \\
    \textit{2021} & \textit{0.1\% (0.0\%)} & \textit{0.6\% (0.1\%)} & \textit{0.3\% (0.1\%)} & \textit{1.0\% (0.1\%)} & \textit{0.1\% (0.0\%)} & \textit{1.6\% (0.2\%)} & \textit{0.1\% (0.0\%)} \\
    \textit{2022} & \textit{0.3\% (0.1\%)} & \textit{1.0\% (0.2\%)} & \textit{0.4\% (0.1\%)} & \textit{1.6\% (0.2\%)} & \textit{0.1\% (0.0\%)} & \textit{2.8\% (0.3\%)} & \textit{0.2\% (0.1\%)} \\
    \hline
    \noalign{\vskip 2pt}
    \multicolumn{8}{l}{\textit{Detection rates (post-ChatGPT)}} \\[2pt]
    2023 & 9.0\% (0.4\%) & 19.3\% (0.5\%) & 10.5\% (0.4\%) & 24.0\% (0.6\%) & 7.4\% (0.4\%) & 27.5\% (0.6\%) & 8.1\% (0.4\%) \\
    2024 & 21.9\% (0.5\%) & 38.0\% (0.6\%) & 24.0\% (0.5\%) & 45.4\% (0.6\%) & 17.8\% (0.5\%) & 40.7\% (0.6\%) & 20.5\% (0.6\%) \\
    2025 & 29.4\% (0.6\%) & 44.9\% (0.7\%) & 32.6\% (0.6\%) & 53.7\% (0.7\%) & 23.2\% (0.6\%) & 34.5\% (0.6\%) & 28.0\% (0.7\%) \\
    \hline\hline
  \end{tabular*}
  \par\medskip
  \parbox{\columnwidth}{%
    \normalfont\emph{Note: Entries report the percentage of essays classified as AI-written, with binomial standard errors in parentheses. Italicized rows correspond to pre-ChatGPT application cycles and are interpreted as false-positive rates. Application year refers to the year in which the application was submitted. Column (7) excludes the changed essay from both the numerator and denominator.}
  }
\end{table}

\end{landscape}

\begin{table}[p]
  \centering
  \normalsize
  \setlength{\tabcolsep}{4pt}
  \renewcommand{\arraystretch}{1.15}
  \caption{Quality Model Dimension Weights}
  \label{tab:quality_weights}
  \begin{minipage}{0.70\columnwidth}
    \centering
    \begin{tabular*}{\linewidth}{@{\extracolsep{\fill}}lcc@{}}
      \hline\hline
      Dimension & Coefficient & 95\% confidence interval \\
      \hline
    Grammar & 0.208 & [0.063, 0.354] \\
    Style & 0.207 & [0.023, 0.391] \\
    Clarity & 0.105 & [-0.132, 0.343] \\
    Prompt responsiveness & 0.479 & [0.014, 0.944] \\
    Voice & 0.226 & [0.107, 0.346] \\
    Public service & 0.213 & [0.059, 0.368] \\
    Leadership & 0.064 & [-0.073, 0.201] \\
      \hline\hline
    \end{tabular*}
    \par\medskip
    \parbox{\linewidth}{%
      \normalfont\emph{Note: Coefficients are estimated from an OLS regression of human quality ratings on the seven LLM-based quality dimensions with rater fixed effects. Brackets report 95\% confidence intervals. The intercept and rater fixed effects are omitted.}
    }
  \end{minipage}
\end{table}

\begin{table}[p]
  \centering
  \begin{minipage}{0.80\columnwidth}
    \centering
    \footnotesize
    \setlength{\tabcolsep}{2pt}
    \renewcommand{\arraystretch}{1.08}
    \caption{Residual Variation in AI Essay Count}
    \label{tab:residuals}
    \begin{tabular*}{\linewidth}{@{\extracolsep{\fill}}lccccc@{}}
      \hline\hline
      \noalign{\vskip 2pt}
       &  & \multicolumn{2}{c}{\makecell[c]{Applicant controls \\[1pt] only}} & \multicolumn{2}{c}{\makecell[c]{Applicant controls \\[1pt] + essay quality}} \\
      \cline{3-4}\cline{5-6}
      \noalign{\vskip 2pt}
      Treatment & Sample & $\widehat{E[V^2]}$ & $R^2$ & $\widehat{E[V^2]}$ & $R^2$ \\[2pt]
      \hline
    \multirow{3}{*}{\makecell[l]{GPTZero: AI Only \\[1pt] High Confidence}} & Pooled & 1.550 & 0.172 & 1.276 & 0.319 \\
     & 2023 & 0.960 & 0.113 & 0.804 & 0.257 \\
     & 2024 & 2.042 & 0.121 & 1.719 & 0.260 \\
    \hline
    \multirow{3}{*}{\makecell[l]{GPTZero: AI + Mixed \\[1pt] High Confidence}} & Pooled & 2.446 & 0.216 & 1.851 & 0.407 \\
     & 2023 & 1.898 & 0.150 & 1.495 & 0.330 \\
     & 2024 & 2.887 & 0.161 & 2.178 & 0.367 \\
    \hline
    \multirow{3}{*}{\makecell[l]{GPTZero: AI Only \\[1pt] Any Confidence}} & Pooled & 1.670 & 0.184 & 1.394 & 0.319 \\
     & 2023 & 1.049 & 0.127 & 0.894 & 0.256 \\
     & 2024 & 2.206 & 0.125 & 1.864 & 0.260 \\
    \hline
    \multirow{3}{*}{\makecell[l]{GPTZero: AI + Mixed \\[1pt] Any Confidence}} & Pooled & 2.959 & 0.226 & 2.243 & 0.414 \\
     & 2023 & 2.374 & 0.171 & 1.905 & 0.335 \\
     & 2024 & 3.430 & 0.161 & 2.541 & 0.378 \\
    \hline
    \multirow{3}{*}{\makecell[l]{Pangram: AI Only}} & Pooled & 1.438 & 0.165 & 1.231 & 0.285 \\
     & 2023 & 0.866 & 0.115 & 0.757 & 0.227 \\
     & 2024 & 1.908 & 0.131 & 1.713 & 0.220 \\
    \hline
    \multirow{3}{*}{\makecell[l]{Pangram: AI + Mixed}} & Pooled & 2.761 & 0.246 & 2.187 & 0.403 \\
     & 2023 & 2.485 & 0.241 & 2.118 & 0.353 \\
     & 2024 & 2.988 & 0.209 & 2.342 & 0.380 \\
    \hline
    \multirow{3}{*}{\makecell[l]{GPTZero: AI Only \\[1pt] Excl. Changed Essay}} & Pooled & 1.019 & 0.148 & 0.853 & 0.286 \\
     & 2023 & 0.605 & 0.102 & 0.517 & 0.233 \\
     & 2024 & 1.352 & 0.101 & 1.170 & 0.222 \\
      \hline\hline
    \end{tabular*}
    \par\medskip
    \parbox{\linewidth}{%
      \normalfont\emph{Note: All specifications include applicant controls; the right panel additionally includes essay-quality measures. Residual variation is the mean squared out-of-fold treatment residual, \(\frac{1}{n}\sum_i(D_i-\widehat{m}(X_i))^2\). Entries are medians across 20 independently seeded fits. Pooled models include an admissions-cycle indicator; year-specific models are estimated separately within each cohort.}
    }
  \end{minipage}
\end{table}

\begin{table}[p]
  \centering
  \begin{minipage}{0.55\columnwidth}
  \centering
  \small
  \setlength{\tabcolsep}{2pt}
  \renewcommand{\arraystretch}{1.08}
  \caption{Pangram AI Usage and Admission}
  \label{tab:admissions_pangram}
  \begin{tabular*}{\linewidth}{@{\extracolsep{\fill}}lccc@{}}
    \hline\hline
    \noalign{\vskip 2pt}
     & \multicolumn{1}{c}{OLS} & \multicolumn{2}{c}{DML} \\
    \cline{2-4}
    \noalign{\vskip 1pt}
     & \multicolumn{1}{c}{(1)} & \multicolumn{1}{c}{(2)} & \multicolumn{1}{c}{(3)} \\[1pt]
    \hline
    \multirow{2}{*}{AI Count} & -0.083*** & -0.016** & -0.022*** \\
     & (0.007) & (0.006) & (0.006) \\
    \hline
    \multirow{2}{*}{AI Count (2023)} & -0.086*** & -0.026* & -0.019 \\
     & (0.014) & (0.011) & (0.012) \\
    \multirow{2}{*}{AI Count (2024)} & -0.082*** & -0.016* & -0.020** \\
     & (0.008) & (0.007) & (0.007) \\
    \hline
    Applicant controls &  & $\checkmark$ & $\checkmark$ \\
    Essay quality &  &  & $\checkmark$ \\
    \hline\hline
  \end{tabular*}
  \par\medskip
  \parbox{\linewidth}{%
    \normalfont\emph{Note: Standard errors are reported in parentheses. The first row reports estimates from models without year interactions; the final two rows report estimates from separate year-specific models. Column (1) reports unadjusted OLS estimates from linear probability models. Columns (2) and (3) report DML estimates using random forest nuisance functions and 5-fold cross-fitting. Across 20 independently seeded fits, coefficients and standard errors are aggregated using DoubleML's median-based repeated-splitting procedure~\citep{chernozhukov_doubledebiased_2018}. \({}^{***}\)~\(p < 0.001\), \({}^{**}\)~\(p < 0.01\), \({}^{*}\)~\(p < 0.05\).}
  }
  \end{minipage}
\end{table}

\begin{table}[p]
  \centering
  \footnotesize
  \setlength{\tabcolsep}{2pt}
  \renewcommand{\arraystretch}{1.08}
  \caption{DML Estimates: Robustness Across Detector Specifications}
  \label{tab:dml_robustness}
  \begin{tabular*}{\columnwidth}{@{\extracolsep{\fill}}lccccc@{}}
    \hline\hline
     & \multicolumn{5}{c}{Applicant controls} \\
    \cline{2-6}
    \noalign{\vskip 2pt}
     & \makecell[c]{GPTZero \\[1pt] AI + Mixed \\[1pt] High Confidence} & \makecell[c]{GPTZero \\[1pt] AI Only \\[1pt] Any Confidence} & \makecell[c]{GPTZero \\[1pt] AI + Mixed \\[1pt] Any Confidence} & \makecell[c]{Pangram \\[1pt] AI + Mixed} & \makecell[c]{GPTZero \\[1pt] AI Only \\[1pt] Excl. Changed Essay} \\[1pt]
     & \multicolumn{1}{c}{(1)} & \multicolumn{1}{c}{(2)} & \multicolumn{1}{c}{(3)} & \multicolumn{1}{c}{(4)} & \multicolumn{1}{c}{(5)} \\
    \hline
    \multirow{2}{*}{Pooled} & -0.015** & -0.015** & -0.014** & -0.012** & -0.017* \\
     & (0.005) & (0.005) & (0.004) & (0.005) & (0.007) \\
    \hline
    \multirow{2}{*}{AI Count (2023)} & -0.026** & -0.017 & -0.018* & -0.013 & -0.028 \\
     & (0.009) & (0.011) & (0.008) & (0.008) & (0.014) \\
    \multirow{2}{*}{AI Count (2024)} & -0.008 & -0.010 & -0.009 & -0.009 & -0.014 \\
     & (0.006) & (0.006) & (0.005) & (0.006) & (0.008) \\
    \hline
     & \multicolumn{5}{c}{Applicant controls + essay quality} \\
    \cline{2-6}
    \noalign{\vskip 2pt}
     & \makecell[c]{GPTZero \\[1pt] AI + Mixed \\[1pt] High Confidence} & \makecell[c]{GPTZero \\[1pt] AI Only \\[1pt] Any Confidence} & \makecell[c]{GPTZero \\[1pt] AI + Mixed \\[1pt] Any Confidence} & \makecell[c]{Pangram \\[1pt] AI + Mixed} & \makecell[c]{GPTZero \\[1pt] AI Only \\[1pt] Excl. Changed Essay} \\[1pt]
     & \multicolumn{1}{c}{(1)} & \multicolumn{1}{c}{(2)} & \multicolumn{1}{c}{(3)} & \multicolumn{1}{c}{(4)} & \multicolumn{1}{c}{(5)} \\
    \hline
    \multirow{2}{*}{Pooled} & -0.027*** & -0.021*** & -0.025*** & -0.021*** & -0.025** \\
     & (0.006) & (0.006) & (0.005) & (0.005) & (0.008) \\
    \hline
    \multirow{2}{*}{AI Count (2023)} & -0.033*** & -0.022 & -0.024** & -0.015 & -0.024 \\
     & (0.010) & (0.012) & (0.009) & (0.008) & (0.016) \\
    \multirow{2}{*}{AI Count (2024)} & -0.020** & -0.018** & -0.022*** & -0.019** & -0.025** \\
     & (0.007) & (0.007) & (0.006) & (0.007) & (0.008) \\
    \hline\hline
  \end{tabular*}
  \par\medskip
  \parbox{\columnwidth}{%
    \normalfont\emph{Note: Standard errors are reported in parentheses. The first row of each panel reports pooled estimates; the final two rows report estimates from separate year-specific models. All models use random forest nuisance functions and 5-fold cross-fitting. Across 20 independently seeded fits, coefficients and standard errors are aggregated using DoubleML's median-based repeated-splitting procedure~\citep{chernozhukov_doubledebiased_2018}. Column (5) excludes the changed essay from treatment construction and from the essay-quality controls in the lower panel. \({}^{***}\)~\(p < 0.001\), \({}^{**}\)~\(p < 0.01\), \({}^{*}\)~\(p < 0.05\).}
  }
\end{table}

\begin{table}[t]
  \centering
  \normalsize
  \setlength{\tabcolsep}{3pt}
  \renewcommand{\arraystretch}{1.08}
  \caption{DML Estimates: Binary Treatment Robustness}
  \label{tab:dml_binary}
  \begin{minipage}{0.90\columnwidth}
    \centering
    \begin{tabular*}{\linewidth}{@{\extracolsep{\fill}}lcccc@{}}
      \hline\hline
       & \multicolumn{2}{c}{Applicant controls} & \multicolumn{2}{c}{\makecell[c]{Applicant controls \\[1pt] + essay quality}} \\
      \cline{2-3}\cline{4-5}
      \noalign{\vskip 2pt}
       & \makecell[c]{GPTZero \\[1pt] Binary} & \makecell[c]{Pangram \\[1pt] Binary} & \makecell[c]{GPTZero \\[1pt] Binary} & \makecell[c]{Pangram \\[1pt] Binary} \\[1pt]
       & \multicolumn{1}{c}{(1)} & \multicolumn{1}{c}{(2)} & \multicolumn{1}{c}{(3)} & \multicolumn{1}{c}{(4)} \\
      \hline
      \multirow{2}{*}{Pooled} & -0.036* & -0.060** & -0.056** & -0.076*** \\
       & (0.018) & (0.019) & (0.019) & (0.020) \\
      \hline
      \multirow{2}{*}{Any AI (2023)} & -0.032 & -0.050 & -0.043 & -0.053 \\
       & (0.031) & (0.032) & (0.032) & (0.034) \\
      \multirow{2}{*}{Any AI (2024)} & -0.035 & -0.049* & -0.061** & -0.070** \\
       & (0.022) & (0.023) & (0.023) & (0.024) \\
      \hline\hline
    \end{tabular*}
    \par\medskip
    \parbox{\linewidth}{%
      \normalfont\emph{Note: The treatment is an indicator for submitting at least one essay classified as AI-written. Columns (1) and (2) include applicant controls; columns (3) and (4) add essay-quality measures. The pooled models include an admissions-cycle indicator, while the 2023 and 2024 estimates are obtained from separate year-specific models. All models use random forest nuisance functions and 5-fold cross-fitting. Across 20 independently seeded fits, coefficients and standard errors are aggregated using DoubleML's median-based repeated-splitting procedure~\citep{chernozhukov_doubledebiased_2018}. Standard errors are reported in parentheses. \({}^{***}\)~\(p < 0.001\), \({}^{**}\)~\(p < 0.01\), \({}^{*}\)~\(p < 0.05\).}
    }
  \end{minipage}
\end{table}

\end{document}